\documentclass[a4paper, 12pt]{amsart}
\usepackage{amsmath, amssymb, amscd, enumerate}

\newtheorem{lem}{Lemma}

\newtheorem{thm}{Theorem}
\newtheorem{prop}{Proposition}
\newtheorem{conj}{Conjecture}

\newcommand{\pf}{{\noindent\bf Proof:\,}}
\makeatletter
\def\@eqnnum{(\thesection.\theequation)}
\makeatother

\begin{document}

\title{ Riemann hypothesis and Quantum Mechanics}

\author{Michel Planat }

\address{Institut FEMTO-ST, CNRS, 32 Avenue de
l'Observatoire, F-25044 Besan\c con, France }

\email{michel.planat@femto-st.fr }

\author{Patrick Sol\'{e} }
\address{Telecom ParisTech, 46 rue Barrault, 75634
Paris Cedex 13, France. }

\email{sole@enst.fr }
\author{Sami Omar }
\address{King Khalid University, Faculty of Sciences, Dept of Math,
9004 Abha, Saudi Arabia also: Faculty of Sciences of Tunis, Dept of Math, 2092 El Manar. Tunis. Tunisia}
\email{Sami.Omar@fst.rnu.tn }
%\subjclass[2000]{Primary 11F11}
\begin{abstract}
In their 1995 paper, Jean-Beno\^{i}t Bost and Alain Connes (BC) constructed a quantum dynamical
system whose partition function is the Riemann zeta function $\zeta(\beta)$, where $\beta$ is an inverse temperature. We
 formulate Riemann hypothesis (RH) as a property of the low temperature Kubo-Martin-Schwinger (KMS) states of this theory.  More precisely, the expectation value
of the BC phase operator can be written as
$$\phi_{\beta}(q)=N_{q-1}^{\beta-1}  \psi_{\beta-1}(N_q),  $$ where $N_q=\prod_{k=1}^qp_k$ is the primorial number of order $q$ and
$ \psi_b $ a generalized Dedekind $\psi$ function depending on one real parameter $b$ as
$$ \psi_b (q)=q \prod_{p \in  \mathcal{P,}p \vert q}\frac{1-1/p^b}{1-1/p}.$$
Fix a large inverse temperature $\beta >2.$ The Riemann hypothesis is then
shown to be equivalent to the inequality
$$ N_q |\phi_\beta (N_q)|\zeta(\beta-1) >e^\gamma \log \log N_q, $$
for $q$ large enough.
Under RH, extra formulas for high temperatures KMS states ($1.5< \beta <2$) are derived.
\end{abstract}


\dedicatory{"Number theory is not pure Mathematics. It is the Physics of the world of Numbers." \\ Alf van der Poorten.}

\maketitle
\section{Introduction}

The Riemann Hypothesis (RH), that describes the non trivial zeroes of Riemann $\zeta$ function, is a Holy
Grail of Mathematics \cite{BCR,L}. Many formulations of RH may be found in the literature \cite{C}. In this paper, we mainly refer to the Nicolas inequality \cite{Nicolas1983} 
\begin{equation}
\frac{N_k}{\varphi(N_k)} >e^\gamma \log \log N_k,
\label{Nicolas}
\end{equation}
where $N_k=\prod_{i=1}^k p_i$ is the primorial of order $k$, $\varphi$ is the Euler totient function and $\gamma \approx 0.577$ is the Euler-Mascheroni constant.

%\begin{itemize}
%\item $\gamma \approx 0.577$ is the Euler Mascheroni constant,
%\item $\varphi$ Euler totient function ,\item
%$N_n=\prod_{k=1}^np_k$ the primorial of order $n,$
%\end{itemize}

The inequality (\ref{Nicolas}) holds for all $k \ge 1$ if RH is true \cite[Th. 2 (a)]{Nicolas1983}.
Conversely, if RH is false, the inequality holds for infinitely
many $k,$ and is violated for infinitely many $k$ \cite[Th. 2
(b)]{Nicolas1983}.\\

The Bost and Connes (BC) system is a quantum dynamical model whose partition function is the Riemann zeta function $\zeta(\beta)$ (with $\beta=\frac{1}{kT}$ the inverse temperature)  and that exhibits a spontaneous symmetry breaking at the pole $\beta=1$. The behaviour  of the system reminds that of a ferromagnet. At a temperature larger than the critical temperature $T_c\approx 10^3 K$, the disorder of the spins dominates and the thermal equilibrium state is unique while for $T<T_C$ the individual spins tend to align to each other, which in the three-dimensional space $\mathbb{R}^3$ of the ferromagnet yields a set of equilibrium thermal states parametrized by the symmetry group $SO(3)$ of rotations in $\mathbb{R}^3$. 
In Bost and Connes approach, the Hamiltonian operator is given by the logarithm of the number operator $N$ as $H_0=\log N$. This contrasts with the ordinary Hamiltonian $H_0=N+\frac{1}{2}$ for the harmonic oscillator. It may be that the logarithmic Hamiltonian is realized in the field of {\it quantum} physiology, where the perception states are proportional to the logarithm of the excitation instead of being proportional to it. For a first account of this process, see \cite{Planat2004,Planat2006}. See also \cite{Schleich2010} for a model of a wave-packet dynamics based on a logarithmic energy spectrum.

The purpose of the paper is two-fold. First, we rewrite the abstract mathematical language of BC theory with, as much as possible, concepts of quantum information processing (QIP). Second, we uncover a relationship between the statistical phase states (KMS states) of  BC theory and Riemann hypothesis (RH). The Riemann hypothesis is shown to be equivalent to an inequality satisfied by the KMS states of the BC theory at a selected low temperature $\beta>2$. A special case (at $\beta=3$) of this inequality is $\psi(N_k)/(N_k \log \log N_k )>e^{\gamma}/\zeta(2)$ for $k>3$, where $\psi(n)=\psi_2(n)$ is the Dedekind $\psi$ function \cite{SolePlanat2010}. It plays a leading role for counting the number of maximal commuting sets of qudits \cite{Planat2011}. Higher temperatures KMS states ($1.5< \beta <2$) require RH to be analyzed with conventional tools such as the prime counting function $\pi(x)$.
Very high temperatures KMS states ($0 <\beta <1.5$) are left as an open problem.
%%%%%%%%%%%%%%%%%%%%%%%%%%%%%%%%%%%%%%%%%%%%%%%%%%%%%%%%%%%%%%%%%%%%%%%%%%%%%%%%%%%%%
%\section{Background on the BC model}
\section{On modular exponentiation: classical and quantum}

Let us start the parallel between the BC theory and QIP with the problem of finding the multiplicative order $r$ in the modular equation
\begin{equation}
a^x=1 \pmod{q}.
\label{modu}
\end{equation}
Given the integer $a$ and a positive integer $q$ with $(a,q)=1$, the order $r=\mbox{ord}_q (a)$ is the smallest positive integer $x$ satisfying (\ref{modu}). It is straigthforward to establish the set of inequalities
\begin{equation}
\mbox{ord}_q (a)\le\lambda(q)\le\varphi(q)\le q-1,
\label{ineq}
\end{equation}
where $\varphi(q)$ is the Euler totient function [it counts the number of irreducible 
fractions $\frac{p}{q}$ with $p<q$] and $\lambda(q)$ is the Carmichael lambda function [it is the smallest positive integer $x$ satisfying (\ref{modu}) for {\it every} integer $a<q$ with $(a,q)=1$] \footnote{The recursive calculation of $\lambda(n)$ is as follows. For a positive integer $k\ge 3$, $\lambda(2^k)=2^{k-2}$. For $p$ an odd prime $\ge 3$ or $0<k\le 2$, $\lambda(p^k)=\varphi(p^k)=(p-1)p^{k-1}$. For $q=\prod_{i=1}^n p_i^{k_i}$ a product of primes, $\lambda(q)=\mbox{lcm}[\lambda(p_i^{k_i})]_i$, where lcm denotes the least common multiple of the numbers in the bracket. }.  Both $\varphi(q)$ and $\lambda(q)$ are {\it universal exponents} that satisfy (\ref{modu}) for any integer $a$ coprime to $q$.

Order-finding is a task that can in principle be performed more efficiently on a quantum computer than on a classical computer. It is a building block of the Shor algorithm for factoring the integers \cite{Nielsen2000}. 

Let $q$ be an $m$-bit integer and $a$ be an integer smaller than $q$ with $(a,q)=1$. The quantum algorithm for order-finding is based on the following unitary operator acting on $m$-bits and such that for any $n<q$
\begin{equation}
\left|n\right\rangle\rightarrow U_a\left|n\right\rangle=\left|an~\mbox{mod}~q\right\rangle.
\label{muop}
\end{equation}
Since $a$ is coprime to $q$, it has an inverse modulo $q$ and $U_a U_a^{\star}=1$ ($U_a$ is unitary). Operator $U_a$ is also multiplicative: $U_a U_b=U _b U_a,~\forall a,b<q$.

The eigenvalues of the operator (\ref{muop}) are in the form $\exp(2i\pi \frac{k}{r})$ and the corresponding eigenvectors are given as a quantum Fourier transform
\begin{equation}
\left|u_k\right\rangle=\frac{1}{r^{1/2}}\sum_{j=0}^{r-1} \exp(\frac{-2i \pi k j}{r})\left|a^j~\mbox{mod}~q\right\rangle.
\label{QFT}
\end{equation}
The efficient implementation of the quantum Fourier transform on a quantum computer allows the efficient estimation of the period $r$ of the function $f(x)= a^x ~\mbox{mod}~q$ in (\ref{modu}), and subsequently the efficient factorization of an integer (in polynomial time instead of the exponential time required on a classical computer).

\section{On the KMS-states of Bost-Connes model}

Bost and Connes theory is a skilfull balance of concepts related to QIP and to quantum statistical physics \cite{Connes1995}. The Bost-Connes $\mathbb{C}^{\star}$ algebra is the cross product \cite{Laca1998}
\begin{equation}
\mathbb{C}_{\mathbb{Q}}=\mathbb{C}^{\star}(\mathbb{Q}/\mathbb{Z}) \rtimes \mathbb{N}^{\star}
\label{cross}
\end{equation}
of the $\mathbb{C}^{\star}$-algebra $\mathbb{C}^{\star}(\mathbb{Q}/\mathbb{Z})$ of additive/phase operators $e_\delta$ acting on qudits $\left|n\right\rangle$ as

\begin{equation}
\left|n\right\rangle\rightarrow e_\delta\left|n\right\rangle =\exp(2i\pi \frac{p}{q})\left|n\right\rangle,~~\delta=\frac{p}{q} \in \mathbb{Q}/\mathbb{Z},
\label{phase}
\end{equation}
by the semigroup $\mathbb{N}^{\star}$ of multiplicative operators $\mu_a$

\begin{equation}
\left|n\right\rangle\rightarrow \mu_a\left|n\right\rangle=\left|an~\mbox{mod}~q\right\rangle,~~a\in \mathbb{N}^{\star}.
\label{newmuop}
\end{equation}
The operator $\mu_a$ is closely related to $U_a$ but acts on qudits $\left|n\right\rangle$.

The BC system is the pair $(\mathbb{C}_{\mathbb{Q}},\sigma_t)$ with entries the algebra $\mathbb{C}_{\mathbb{Q}}$ and a quantum statistical flow $\sigma_t$.

In standard quantum statistical mechanics, given an observable Hermitian operator $M$ and a Hamiltonian $H_0$, one has the evolution $\sigma_t(M)$ versus time $t$ as
\begin{equation}
\sigma_t(M)=\exp(itH_0)M\exp(-itH_0),
\end{equation}
and the expectation value of $M$ at the inverse temperature $\beta=\frac{1}{kT}$ is the unique Gibbs state
\begin{equation}
\mbox{Gibbs}(M)=\mbox{Trace}(M \exp(-\beta H_0))/\mbox{Trace}(\exp(-\beta H_0).
\end{equation}
For the more general case of an algebra of observables $A$, the Gibbs state is replaced by the so-called Kubo-Martin-Schwinger (or KMS$_\beta$ state). One introduces a $1$-parameter group $\sigma_t$ of automorphisms of $A$ so that for any $t \in \mathbb{R}$ and $x \in A$ 
\begin{equation}
\sigma_t(x)=\exp(itH_0)x\exp(-itH_0).
\label{flow}
\end{equation}
For the algebra $A$, the equilibrium state remains unique only if some conditions regarding the evolution of the operator $x \in A$ are satisfied: the KMS states $\phi_{\beta}$ on A have to fulfill the KMS$_\beta$ condition with respect to $\sigma_t$, namely that for any $x,y \in A$, there exists a bounded holomorphic function $F_{x,y}(z)$, $0\le \Im z \le \beta$ such that
\begin{equation}
F_{x,y}(t)=\phi_{\beta}(x\sigma_t(y))~\mbox{and}~F_{x,y}(t+i \beta)=\phi_{\beta}(\sigma_t(y)x),~~\forall t \in \mathbb{R}.
\end{equation}

\subsection*{The KMS flow of BC system}

The Hamiltonian on the BC system is defined from its action on the qudits $\left|n\right\rangle$ by
\begin{equation}
H_0\left|n\right\rangle=\ln n \left|n\right\rangle.
\label{Hami}
\end{equation}
Using the relations $\exp(-\beta H_0)\left|n\right\rangle=\exp(-\beta \ln n)\left|n\right\rangle=n^{-\beta}\left|n\right\rangle$, it follows that the partition function of the model at the inverse temperature $\beta$ is
\begin{equation}
\mbox{Trace}(\exp(-\beta H_0))=\sum_{n=1}^{\infty}n^{-\beta}=\zeta(\beta),~~\Re(\beta)>1.
\label{parti}
\end{equation}
Applying (\ref{flow}) to the Hamiltonian (\ref{Hami}) one obtains
\begin{equation}
\sigma_t(\mu_a)=a^{it}\mu_a~\mbox{and}~\sigma_t(e_\delta)=e_\delta~\mbox{for}~\mbox{any}~ a \in \mathbb{N}^{\star},~\delta \in \mathbb{Q}/\mathbb{Z} ~\mbox{and}~t \in\mathbb{R}.
\end{equation}
It can be shown \cite{Connes1995,Laca1998} that in the high temperature regime $0<\beta \le 1$ there is a unique KMS$_\beta$ state of the quantum statistical system ($\mathbb{C}_{\mathbb{Q}},\sigma_t$). It corresponds to the partition function (\ref{parti}). In the low temperature regime $\beta >1$, the KMS$_\beta$ is no longer unique. The symmetry group $G$ of the quantum statistical system is a subgroup of the automorphisms of ($\mathbb{C}_{\mathbb{Q}},\sigma_t$) which commutes with $\sigma_t$, i.e.
\begin{equation}
w\circ \sigma_t=\sigma_t \circ w ~\mbox{for}~\mbox{any}~w\in G~\mbox{and}~ t \in \mathbb{R},
\label{comm}
\end{equation}
where $\circ$ is the composition law. The symmetry group $G$ permutes a family of extremal KMS$_\beta$ states generating the possible states of the system after phase transition. The symmetry group of the algebra $(\mu_a,e_\delta)$ endowed with the Hamiltonian $H_0$ in (\ref{Hami}) is the Galois group $G=\mbox{Gal}(\mathbb{Q}^{\mbox{cycl}}/\mathbb{Q})$ of the cyclotomic extension $\mathbb{Q}^{\mbox{cycl}}$ of $\mathbb{Q}$. The field $\mathbb{Q}^{\mbox{cycl}}$ is generated in $\mathbb{Q}$ by all the roots of unity. The Galois group $G$ is $\mbox{aut}(\mathbb{Q}^{\mbox{cycl}})$. It is isomorphic to $\mathbb{Z}_q^\star$.
% At this stage, one may remind that the Carmichael lambda function $\lambda_q$ is the exponent in $\mathbb{Z}_q^\star$.  

At any temperature $\beta< \infty$, the restriction of the KMS$_\beta$ state to $\mathbb{C}^{\star}(\mathbb{Q}/\mathbb{Z})$ is given by
\begin{equation}
\phi_\beta(q)=q^{-\beta}\prod_{p\in\mathcal{P},p|q}\frac{1-p^{\beta-1}}{1-p^{-1}}
\label{KMS}
\end{equation}
and for $\beta=\infty$ it is
\begin{equation}
\phi_\infty(q)=\mu(q)/\varphi(q),
\label{KMSinf}
\end{equation}
where $\mu$ is the M\"{o}bius function.

The KMS state (\ref{KMS}) may be easily rewritten as $\phi_{\beta}(q)=N_{q-1}^{\beta -1}(q) \psi_{\beta-1}(q)$, where $N_q=\prod_{k=1}^q p_k$ is a primorial number and $\psi_b(q)$ is defined in the abstract.

\section{The Bost-Connes KMS states and Riemann hypothesis}

In section(\ref{subs1}) one formulates Riemann hypothesis as a property of the KMS states (\ref{KMS}) with the inequality
\begin{equation}
\epsilon_{\beta}(q)=\frac{N_q|\phi_\beta (N_q)|}{\log \log N_q} -\frac{e^\gamma}{\zeta(\beta-1)}>0 ~\mbox{when}~\beta>2. 
\label{KMSineq}
\end{equation}
 The special case $\beta=3$ corresponds to our recent paper \cite{SolePlanat2010}. In section (\ref{subs2}), new conditional formulas are established.
  
A concise account of numerical values at selected temperature are in table 1.

\begin{table}[ht]
\begin{center}
\small
\begin{tabular}{|r|r|r|r|r|}
\hline
$q$ & $10$ & $10^2$ & $10^3$ & $10^4$ \\
\hline
$N_q$ & $6.4 \times 10^{10}$ & $4.7 \times 10^{219}$ & $6.7 \times 10^{3392}$ & $9.1 \times 10^{45336}$\\
\hline
$\epsilon_{2.1}$ & $0.25$ & $0.093$ & $0.051$ & $0.031$ \\
\hline
$\epsilon_{3}$ & $0.16$ & $0.018$ & $3.0 \times 10^{-3}$ & $6.1 \times 10^{-4}$ \\
\hline
$\epsilon_{10}$ & $0.25$ & $0.028$ & $4.9 \times 10^{-3}$ & $1.0 \times 10^{-3}$ \\
\hline
\end{tabular}
\label{table1} 
\caption{A check of KMS inequality (\ref{KMSineq}).}
\end{center}
\end{table}

%%%%%%%%%%%%%%%%%%%%%%%%%%%%%%%%%%%%%%%%%%%%%%%%%%%%%%%%%%%%%%%%%%%%%%%%%%%%%%%%%%%%%
\subsection{Low temperature KMS states}
\label{subs1}
To simplify the notation, we define the ratio $R_b(n):=\frac{\psi_b(n)}{n\log\log n}.$ The motivation
for this strange ratio will become apparent in the following application of Mertens formula \cite[Th. 429]{HW}.

{\prop \label{mertens}  Pick a real number $b>1.$ For $n$ going to $\infty,$ we have $$\lim R_b(N_n)=\frac{e^\gamma}{\zeta(b)}.$$}

\pf
By the definition of $\psi_b(n)$
we can combine the Eulerian product for $\zeta(b)$ with the Mertens formula
$$\prod_{p\le x}(1-1/p)^{-1}\sim e^\gamma \log(x).$$

to obtain
	
$$\psi_b(N_n)\sim \frac{e^\gamma}{\zeta(b)}\log(p_n),$$
Now the Prime Number Theorem \cite[ Th. 6, Th. 420]{HW} shows that $x\sim \theta(x)$ for $x$ large
where $\theta(x)$ stands for Chebyshev's first summatory function:
$$\theta(x)= \sum_{p\le x }\log p.$$
This implies that, upon taking $x = p_n$ we have
$p_n \sim \theta(p_n) = \log(N_n).$
The result follows. \qed
%This motivates the search for explicit upper bounds on R(N

{\prop \label{trick} Pick a real number $b>1.$ For $n\ge 2$ we have $$n^2 >\varphi(n)\psi_b(n)\ge \frac{n^2}{\zeta(b)} .$$}
\pf
The first inequality follows at once upon writing
$$\frac{\varphi(n)\psi_b(n)}{n^2}=\prod_{p \vert n} (1-1/p^b),$$
a product of finitely many terms $<1.$
Notice for the second inequality that
$$\frac{\varphi(n)\psi_b(n)}{n^2}=\prod_{p \vert n} (1-1/p^b)\ge \prod_{p } (1-1/p^b),$$

an infinite product that is the inverse of the Eulerian product for $\zeta(b).$

\qed

The next result is a far reaching consequence of Nicolas study of the small values of Euler $\varphi$ function. 

{\thm \label{lower} Under RH the ratio $R_b(N_n)$ is $> \frac{e^\gamma}{\zeta(b)}$ for $n \ge 3$. If RH is false, this is still true for infinitely many $n.$ }

\pf

Follows by Proposition \ref{trick}, combined with \cite[Theorem 2]{Nicolas1983}.
\qed

In view of this result and of numerical experiments the natural conjecture is 

{\conj \label{natural} Fix a real number $b>1.$ For all  $n \ge 3$ we have  $R_b(N_n)> \frac{e^\gamma}{\zeta(b)}.$}

The main result of this note is the following.
{\thm \label{main}  Conjecture \ref{natural} is equivalent to RH.}

\pf
If RH is true we refer to the first statement of Theorem \ref{lower}. If RH is false we consider the function
$$g(x):=\frac{e^\gamma}{\zeta(b)}\log\theta(x)\frac{\prod_{p\le x}(1-1/p)}{\prod_{p\le x}(1-1/p^b)}.$$

 Observing that $\log \theta(p_n)=\log N_n,$ we see that
 $g(p_n) <1$ is equivalent to $R_b(N_n)> \frac{e^\gamma}{\zeta(b)}.$ We need to check that for $x$ large enough $g(x)$ can be $>1$ or equivalently 
$\log g(x) >0.$ Using  \cite[Lemma 6.4]{CLMS}, we obtain, upon writing
$$-\log \zeta(b)=\sum_{p\le x}\log(1-1/p^b)+\sum_{p> x}\log(1-1/p^b), $$
the bound $\log g(x) \ge \log f(x)-b/x,$ where $f$ is the function introduced in \cite[Theorem 3]{Nicolas1983}, that is

$$f(x):={e^\gamma}\log\theta(x)\prod_{p\le x}(1-1/p).$$

We know by \cite[Theorem 3 (c)]{Nicolas1983} that, if RH is false, there is a constant $0<c<1$ such that $\limsup x^{-c}f(x) >0.$

Since for $x$ large $b/x <<x^{-c},$ the result follows.
\qed

\subsection{ High temperature KMS states}
\label{subs2}

To investigate high temperature KMS states at large dimensions $n$, we will need the following lemmas. Recall $\pi(x)$ the prime counting function

$$\pi(x)=\sum_{p\le x}1.$$

{\lem Let $0 <b<1.$ We have $$\sum_{p\le x}\frac{1}{p^b}=\frac{\pi(x)}{x^{b}}+bJ_b(x),$$

where $J_b(x)=\int_2^x\frac{\pi(t)dt}{t^{1+b}}.$
}

\pf Write the above sum as a Stieltjes integral, and integrate by parts.\qed

Recall the Prime Number Theorem with rest from 1896 \cite{HW},

$$\pi(x)=Li(x)+O(x \exp(-c\sqrt{\log x})),$$

for some constant $c>0,$
as well as the Prime Number Theorem with rest under RH (\cite{Koch1901}).
$$\pi(x)=Li(x)+O(\sqrt{ x }\log x ).$$
Plugging these estimates into the definition of $J_b$ we obtain
{\lem Let $0 <b<1.$ We have, unconditionally, $$J_b(x)=\int_2^x\frac{Li(t)dt}{t^{1+b}}+O(\int_2^x\frac{\exp(-c\sqrt{\log t}) dt}{t^b }).$$

and, under RH
$$J_b(x)=\int_2^x\frac{Li(t)dt}{t^{1+b}}+O(\int_2^x \frac{\log t dt}{t^{b+0.5}}).$$
}

The preceding result shows that the assumption $b>1/2$ is necessary to have asymptotic results to the precision $O(1),$ since we need the integral in the error term to converge.
{\lem Let $0.5<b<1.$ We have, under RH
$$\sum_{p\le x}\frac{1}{p^b}=\frac{Li(x)}{x^b}+bI_b(x)+O(1),$$  where
$$I_b(x)=\int_2^x\frac{Li(t)dt}{t^{1+b}}.$$
}

\pf Combine Lemma 1, Koch's paper \cite{Koch1901} and Lemma 2.\qed

{\thm Let $0.5<b<1.$ We have, under RH, $$\sum_{p\le x}\frac{1}{p^b}=\int_2^x\frac{ dt}{t^b \log t}+O(1).$$}

\pf Use lemma 3 and integrate $I_b$ by parts.\qed

If $0.5<b<1,$ denote by $C_b$ the sum of the convergent series
$$C_b=\sum_{p }(\log( 1-1/p^b )+1/p^b).$$

Denote by $A_b(x)$ the partial sum 
$$A_b(x)=\sum_{p \le x}\log( 1-1/p^b ).$$
Denote by $B_b(x)$ the Bertrand integral $B_b(x)=\int_2^x\frac{ dt}{t^b \log t}.$
{\thm \label{asymp} Under RH and for $0.5<b<1,$ there is constant $K_b>0,$ such that, for large $n,$ we have
$$ \psi_b(N_n) \sim K_b\log p_n \exp(-B_b(p_n)).$$}

\pf Note that the definition of $A_b(x)$ entails 
$$A_b(x)=C_b+o(1)- \sum_{p \le x} 1/p^b.$$
Combine Theorem 1 with the above and with Mertens formula
$$\prod_{p\le x}(1-1/p)^{-1}\sim e^\gamma \log x.$$
The result follows upon putting $x=p_n.$
\qed

We now give a lower bound on $\psi_b(N_n).$
{\thm Under RH and for $0.5<b<1,$ for every $\epsilon>0,$ there is an $N$ such that
$n\ge N$ yields
$$\psi_b(n)>K_b e^\gamma (1-\epsilon) N_n \log_2^2N_n \exp(-B_b(p_n)).$$
}

\pf  First, note that
$$\frac{\varphi(N_n)}{N_n}\psi_b(N_n)=\prod_{p\le p_n}(1-1/p^b),$$

a divergent product bounded below by
$$K_b(1-\epsilon) \log_2N_n \exp(-B_b(p_n)),$$ for $n$ large enough, by Theorem \ref{asymp}.

Under RH Nicolas inequality 
$$\frac{N_n}{\varphi(N_n)}> e^\gamma N_n \log_2 N_n,$$ holds for all $n.$ The result follows.
\qed

\section{Conclusion and open problems}
In the present paper, we established the new criterion (\ref{KMSineq}) (theorem 2) for Riemann hypothesis from the low temperature KMS states of the Bost-Connes model. 
%In the present paper we have derived a criterion for Riemann Hypothesis in the context of the Bost Connes quantum statistical model for prime numbers.
Our vision is related to but distinct from Connes vision  of the zeroes of $\zeta$ as an absorption spectrum \cite{Connes99}.
 In our approach the zeroes lurk in the backstage, and partial Eulerian products appear on the forefront. While the low temperatures KMS states ($\beta>2$) lend themselves to an elegant analysis leading to infinitely many equivalence criterion with RH, high temperatures ($2>\beta>3/2$) forced us to assume RH to establish the asymptotic formulas in theorems 4 and 5.
 % derive asymptotics and lower bounds.

Future work should focus on the physical understanding of the lower bound on KMS states (equivalent to RH) we established. Another direction of inquiry  is the study of such a lower bound attached to a quantum statistical system whose partition function is the Dedekind zeta function \cite{Cohen1999}. As a finer point, we mention very high temperatures KMS states 
( $\beta<3/2$), which are challenging to analyze even under RH.

{\bf Acknowledgement:} The second and third author are grateful to the Middle East Center of Algebra and its Applications (MECAA) at King Abdulaziz University for its warm and fruitful hospitality.
 
%%%%%%%%%%%%%%%%%%%%%%%%%%%%%%%%%%%%%%%%%%%%%%%%%%%%%%%%%%%%%%%%%%%%%%%%%%%%%%%%%%%%%

\bibliographystyle{amsplain}

\begin{thebibliography}{10}
%\bibitem{A}Tom M. Apostol,{ \it Introduction to analytic number
%theory.} Undergraduate Texts in Mathematics. Springer-Verlag, New
%York-Heidelberg, 1976.
\bibitem{BCR}
Borwein P~B, Choi S, Rooney B and Weirathmueller A 2008 {\it The Riemann hypothesis: a
resource for the afficionado and virtuoso alike} (Springer, Berlin).


% Peter B. Borwein, Stephen Choi, Brendan
%Rooney, Andrea Weirathmueller, { \it The Riemann hypothesis: a
%resource for the afficionado and virtuoso alike  }  Canadian Math
%Soc., 2008.

\bibitem{Connes1995}
Bost J~C and Connes A 1995 Hecke algebras, type III factors and phase transitions with spontaneous symmetry breaking in number theory {\it Sel. Math.} {\bf 1} 411-457.

\bibitem{Connes99}
Connes A 1999 Trace formula in noncommutative geometry and the zeros of the Riemann zeta function {\it Sel. Math.} {\bf 5} 29--106.

\bibitem{Cohen1999}
Cohen P~B 1999 Quantum statistical mechanics and number theory {\it Contemp. Math.} {\bf 241} 121--128.

\bibitem{C}
Conrey B~J 2003 The Riemann hypothesis {\it Notices Amer. Math. Soc.} {\bf 50} 341--353.

\bibitem{CLMS}
Choie Y, Lichiardopol N, Moree P and Sol\'e P 2007 On Robin's criterion for the Riemann hypothesis {\it J. Th\'eor. Nombres Bordeaux} {\bf 19} 357--372.

% Choie, YoungJu, Lichiardopol, Nicolas, Moree, Pieter, Sol\'e, Patrick
%On Robin's criterion for the Riemann hypothesis,
%J. Th\'eor. Nombres Bordeaux 19 (2007), no. 2, 357--372.

\bibitem{HW}
 Hardy G~H and Wright E~M 1979 {\it An introduction to the theory of numbers} (Oxford Press, Oxford).

\bibitem{I}
Ingham A~E 1990 {\it The distribution of prime numbers} (Cambridge University Press, Cambridge).

%\bibitem{J}Rafael Jakimczuk, Asymptotic Expansions for $1/p_n$ and $\log p_n/p_n,$ International Mathematical Forum, 5, 2010, no. 57, 2817--2834
\bibitem{L}
Lachaud G 2005 {\it L'hypoth\`ese de Riemann : le Graal des math\'ematiciens}
(La Recherche Hors-S\'erie no. 20).

%\bibitem{N} Jean-Louis Nicolas, Petites valeurs de la fonction d'Euler. J. Number Theory  17  (1983),  no. 3, 375--388.
%\bibitem{SP}Patrick Sol\'e, Michel Planat, Extreme values of the
%Dedekind $\Psi$ function, {\tt arXiv:1011.1825v1 [math.NT]}.
\bibitem{Nielsen2000}
Nielsen A~N and Chuang I~L 2000 {\it Quantum computation and quantum information} (Cambridge University Press, Cambridge).

\bibitem{Planat2004}
Planat M 2004 On the cyclotomic quantum algebra of time perception {\it Neuroquantology} {\bf 2} 292-308.

\bibitem{Planat2006}
Planat M 2006 Huyghens, Bohr, Riemann and Galois: phase-locking {\it Int. J. Mod. Phys. B} {\bf 20} 1833-1850.

\bibitem{SolePlanat2010}
Sol\'e P and Planat M 2010 Extreme values of the Dedekind $\psi$ function. {\it J. Comb. Numb. Theor.} (accepted). Preprint 1011.1825 [math.NT].

\bibitem{Nicolas1983}
Nicolas J~L 1983 Petites valeurs de la fonction d'Euler {\it J. Number Theory} {\bf 17} 375-388.

\bibitem{Koch1901}
von Koch H (1901) Sur la distribution des nombres premiers {\it Acta Mathematica} {\bf 24} 159--182.

\bibitem{Laca1998}
Laca M 1998 Semigroup of *-endomorphisms, Dirichlet series and phase transitions {\it J. Funct. Anal.} {\bf 152} 330-378.

\bibitem{Schleich2010}
Mack R, Dahl J~P, Moya-Cessa H, Strunz W~T, Walser R and Schleich W~P 2010 Riemann zeta function from wave-packet dynamics {\it Phys. Rev. } {\bf A82} 032119.

\bibitem{Planat2011}
Planat M 2011 Pauli graphs when the Hilbert space dimension contains a square: why the Dedekind psi function? {\it J. Phys. A: Math. Theor.} {\bf 44} 045301-16.


\end{thebibliography}

\end{document}